\pdfoutput=1
\RequirePackage{ifpdf}
\ifpdf 
\documentclass[pdftex]{sigma}
\else
\documentclass{sigma}
\fi

\begin{document}

\allowdisplaybreaks

\renewcommand{\thefootnote}{$\star$}

\renewcommand{\PaperNumber}{063}

\FirstPageHeading

\ShortArticleName{Singular Isotonic Oscillator, Supersymmetry and Superintegrability}

\ArticleName{Singular Isotonic Oscillator, Supersymmetry\\ and Superintegrability\footnote{This
paper is a contribution to the Special Issue ``Superintegrability, Exact Solvability, and Special Functions''. The full collection is available at \href{http://www.emis.de/journals/SIGMA/SESSF2012.html}{http://www.emis.de/journals/SIGMA/SESSF2012.html}}}

\Author{Ian MARQUETTE}

\AuthorNameForHeading{I.~Marquette}

\Address{School of Mathematics and Physics, The University of Queensland,\\ Brisbane, QLD 4072, Australia}
\Email{\href{mailto:i.marquette@uq.edu.au}{i.marquette@uq.edu.au}}

\ArticleDates{Received July 20, 2012, in f\/inal form September 14, 2012; Published online September 19, 2012}

\Abstract{In the case of a one-dimensional nonsingular Hamiltonian $H$ and a singular supersymmetric partner $H_{a}$, the Darboux and factorization relations of supersymmetric quantum mechanics can be only formal relations. It was shown how we can construct an adequate partner by using inf\/inite barriers placed where are located the singularities on the real axis and recover isospectrality. This method was applied to superpartners of the harmonic oscillator with one singularity. In this paper, we apply this method to the singular isotonic oscillator with two singularities on the real axis. We also applied these results to four 2D superintegrable systems with second and third-order integrals of motion obtained by Gravel for which polynomial algebras approach does not allow to obtain the energy spectrum of square integrable wavefunctions. We obtain solutions involving parabolic cylinder functions.}

\Keywords{supersymmetric quantum mechanics; superintegrability; isotonic oscillator; polynomial algebra; special functions}

\Classification{81R15; 81R12; 81R50}

\renewcommand{\thefootnote}{\arabic{footnote}}
\setcounter{footnote}{0}

\vspace{-2mm}

\section{Introduction}
In recent years, many papers \cite{Agb,Ber1,Ber2, Car,Fel,Grad,Hal,Kra,Ses} were devoted to a nonsingular isotonic oscillator and its various generalizations. It is also referred in literature as CPRS system and also often written using the following parameter $\omega=\frac{\hbar}{2a^{2}}$ or taking $\hbar=1$. This quantum system is exactly solvable and given by the following equation (i.e.\ with the parameter given by $a=ia_{0}$, $a_{0}\in \mathbb{R}$)
\begin{equation}
H=\frac{P_{x}^{2}}{2} + \hbar^{2}\left(
\frac{x^{2}}{8a^{4}} +
\frac{1}{(x-a)^{2}}+\frac{1}{(x+a)^{2}}\right) . \label{niosc}
\end{equation}
Let us also def\/ine the singular isotonic oscillator that correspond to the Hamiltonian given by the equation~\eqref{niosc} with alternatively the parameter def\/ined as $a \in \mathbb{R}$ and the singularities are on the real axis. Let us mention that in the nonsingular case the solutions in terms of series was obtained~\cite{Car} that involve polynomials that are linear combinations of Hermite polynomials. Moreover, it was shown by J.~Sesma~\cite{Ses} that the corresponding Schr\"odinger equation can be transformed into the conf\/luent Heun equation~\cite{Sla2, Sla1} and studied quasi-exactly solvable analogs. In addition, the CPRS system was related to position-dependent ef\/fective mass Schr\"odinger equations~(PDMS)~\cite{Kra} and it was shown recently, using Bethe ansatz method~\cite{Agb}, that this system possesses also a hidden $sl(2)$ algebraic structure.

Let us mention that the nonsingular isotonic oscillator was obtained and studied in earlier works of Spiridonov and Veselov in the context of the dressing chain method \cite{Spi, Ves} and that some results presented in~\cite{Fel} were obtained in the papers~\cite{Sam, Tka}. This system is also a~parti\-cu\-lar case of a system \cite{And,Dav2,Dav1,Mar2,Mar1} involving the fourth Painlev\'e transcendent~\cite{Inc} and related to third-order supersymmetric quantum mechanics (SUSYQM). Recently,
using a~modif\/ied factorization a family that include this system was also obtained~\cite{Fel}.

The nonsingular isotonic oscillator also appears in context of 2D quantum superintegrable systems with a second and a third-order integrals of motion with separation of variables in Cartesian coordinates classif\/ied by S.~Gravel~\cite{Gra}. Four of these systems that possess more integrals than degrees of freedom, contain the nonsingular isotonic oscillator. These four superintegrable systems were studied \cite{Mar4, Mar3} from the point of view of SUSYQM \cite{Gen, Jun,Wit} and polynomial algebras. The square integrable wavefunctions and the energy spectrum were obtained. Furthermore, the superintegrability property does not depend on the nature of the singularity and thus the 2D superintegrable systems remain superintegrable when we take $a \in \mathbb{R}$ and thus these 2D superintegrable systems also admit two types, i.e.\ one singular and one regular type. The singular version of these 2D superintegrable Hamiltonian remain to be solved.

Let us mention, that the singular isotonic oscillator (i.e.\ equation~\eqref{niosc} with $a \in \mathbb{R}$)
belongs to the class of 1D Hamiltonian constructed by Robnik \cite{Rob} from SUSYQM with higher excited states to construct the superpotential. It was recognized that the factorisation and interwining (or Darboux) relations of SUSYQM are only formal.

It was observed by many authors that for systems with singularities on the real axis related to a regular superpartner, SUSYQM does not allow \cite{Cas,Das,Jev,Marq,Zno1,Zno2} to relate the wavefunctions and the energy spectrum of the superpartners using supercharges. Supersymmetry and it algebraic equations (factorization, Darboux or intertwining) are formal and do not take into account boundaries or singularities. A method to reobtain the isospectrality property, i.e.\ relate the wavefunctions and the energy spectrum for a singular Hamiltonian and its regular superpartner, consists to use inf\/inite barriers to modify the regular Hamiltonian \cite{Marq}. However, this idea was only  applied to systems with only one singularity on the real axis. The study of Hamiltonians with many singularities using this approach is an unexplored subject.

The purpose of this paper is to apply this approach to the singular isotonic oscillator which has 2 singularities of the real axis and also use these results to obtain the energy spectrum and wavefunctions for four 2D singular superintegrable systems with a second and third-order integrals that remains to be solved.

Let us present the organization of the paper. In Section~\ref{section2}, we recall results obtained for the constrained harmonic oscillator. In Section~\ref{section3}, present def\/inition and results concerning supersymmetric quantum mechanics and discussed how we can construct the adequate superpartner for singular and nonsingular superpartners. In Section~\ref{section4}, we apply the results of Section~\ref{section2} and the method of Marquez, Negro and Nieto~\cite{Marq} to the singular isotonic oscillator. We obtain the wavefunctions and square integrable wavefunctions. In Section~\ref{section5}, we discuss how the polynomial algebra obtained for four 2D singular superintegrable systems with a second and a third integrals of motion that contain the singular isotonic oscillator is only formal and related to SUSYQM. We apply the results of Section~\ref{section4} to solve these systems and obtain the corresponding energy spectrum and wavefunctions in terms of the parabolic cylinder functions.

\section{Constrained harmonic oscillator}\label{section2}

Let us recall known results concerning the 1D unconstrained and constrained harmonic oscillator (i.e.\ with one or two symmetric inf\/inite barriers). The unconstrained harmonic oscillator that we denote by $V_{\rm I}$ is given by the following well known Hamiltonian
\begin{equation} \label{hamil}
H\psi=\left(-\frac{\hbar^{2}}{2}\frac{d^{2}}{dx^{2}}+V_{\rm I}(x)\right)\psi(z)=E\psi(x),  \qquad
V_{\rm I}(x)=\frac{\omega^{2}x^{2}}{2},\qquad -\infty < x < \infty .
\end{equation}
The energy spectrum and the wavefunctions are also well known
\begin{equation*}
E_{n}=\left(n+\frac{1}{2}\right)\hbar\omega,\qquad \psi_{n}=\left(\sqrt{\pi}n!\right)^{-\frac{1}{2}}h_{n}\left(\left(\frac{2\omega}{\hbar}\right)^{\frac{1}{2}}x\right), 
\end{equation*}
where $h_{n}$ are the Hermite polynomials.

We can def\/ined on $[B,\infty)$ and $[-B,B]$ respectively the constrained harmonic oscillators $V_{\rm II}$ and $V_{\rm III}$ \cite{Dea,Mei}:
\begin{gather}\label{vII}
V_{\rm II}(x)=
\begin{cases}
\dfrac{\omega^{2}x^{2}}{2},&   x \geq B,\quad B>0 ,\\
\infty, &  x < B, \quad B \in \mathbb{R},
\end{cases}
\\
V_{\rm III}(x)=
\begin{cases}
\dfrac{\omega^{2}x^{2}}{2}, &   -B \leq x \leq B, \quad B>0 , \\
\infty,& x < - B, \quad x > B ,\quad B \in \mathbb{R}.
\end{cases}\label{vIII}
\end{gather}

We consider also the following transformation
\begin{equation*}
z=\sqrt{\frac{2\omega}{\hbar}}x,\qquad \epsilon=-\frac{E}{\hbar\omega},\qquad b=\sqrt{\frac{2\omega}{\hbar}}B,\qquad \psi(x)=y(z),  
\end{equation*}
and the Schr\"odinger equation corresponding to the non vanishing part of these potentials $V_{\rm II}$ and $V_{\rm III}$ thus become the Weber dif\/ferential equation
\begin{equation}
\frac{d^{2}y(z)}{dz^{2}}-\left(\frac{z^{2}}{4}+\epsilon\right)y(z)=0.  \label{equa}
\end{equation}

\subsection[Constrained harmonic oscillator $V_{\rm II}$]{Constrained harmonic oscillator $\boldsymbol{V_{\rm II}}$}

For the quantum system $V_{\rm II}$ given by equation~\eqref{vII} we have the following constraints on the solution $y(z)$ of the  equation~\eqref{equa} \cite{Dea,Marq,Mei}:
\begin{enumerate}\itemsep=0pt
\item[$(i)$] $y(z)$ must be square integrable,

\item[$(ii)$] $y(z)$ must be continuous on $\mathbb{R}$, i.e.\ $y(z)\rightarrow 0$ when $ x \rightarrow b^{+}$.
\end{enumerate}

The equation~\eqref{equa} can be solved in terms of the parabolic cylinder functions \cite{Abr, Whi}
\begin{gather*}
y_{1}(\epsilon,z)=e^{-\frac{1}{4}z^{2}} \,{}_1 F_1 \left(\frac{1}{2}\epsilon +\frac{1}{4},\frac{1}{2},\frac{1}{2}z^{2}\right), 
\qquad
y_{2}(\epsilon,z)=e^{-\frac{1}{4}z^{2}} \,{}_1 F_1\left(\frac{1}{2}\epsilon +\frac{3}{4},\frac{3}{2},\frac{1}{2}z^{2}\right),  
\end{gather*}
where ${}_1 F_1$ is the conf\/luent hypergeometric function also related to the Whittaker function.
We can also introduce the following and more appropriate functions in the case of this constrained harmonic oscillator \cite{Abr, Whi} $U(\epsilon,z)$ and $V(\epsilon,z)$ that are combinations of $y_{1}(\epsilon,z)$ and $y_{2}(\epsilon,z)$ and given by
\begin{gather*}
U(\epsilon,z)=Y_{1}\cos\left(\pi\left(\frac{1}{4}+\frac{1}{2}\epsilon\right)\right)-Y_{2}\sin\left(\pi\left(\frac{1}{4}+\frac{1}{2}\epsilon\right)\right),  
\\
V(\epsilon,z)=\frac{1}{\Gamma\left(\frac{1}{2}-\epsilon\right)}\left(Y_{1}\sin\left(\pi\left(\frac{1}{4}+\frac{1}{2}\epsilon\right)\right)
+Y_{2}\cos\left(\pi\left(\frac{1}{4}+\frac{1}{2}\epsilon\right)\right)\right),  
\end{gather*}
where
\begin{equation*}
Y_{1}(\epsilon,z)=\frac{\pi^{\frac{1}{2}}\sec\left(\pi\left(\frac{1}{4}+\frac{1}{2}\epsilon\right)\right)}
{2^{\frac{1}{2}\epsilon+\frac{1}{4}}\Gamma\left(\frac{3}{4}+\frac{1}{2}\epsilon\right)}y_{1}(\epsilon,z),\qquad   Y_{2}(\epsilon,z)=\frac{\pi^{\frac{1}{2}}\operatorname{cosec}\left(\pi\left(\frac{1}{4}+\frac{1}{2}\epsilon\right)\right)}
{2^{\frac{1}{2}\epsilon-\frac{1}{4}}\Gamma\left(\frac{1}{4}+\frac{1}{2}\epsilon\right)}y_{2}(\epsilon,z). 
\end{equation*}
The asymptotic behavior of the functions $U(\epsilon,z)$ and $V(\epsilon,z)$ is given by the following expressions
\begin{gather*}
U(\epsilon,z)\sim \frac{e^{-\frac{z^{2}}{4}}}{z^{\epsilon+\frac{1}{2}}}
\left(1-\frac{\left(\epsilon+\frac{1}{2}\right)\left(\epsilon+\frac{3}{2}\right)}{2z^{2}}+\cdots \right) , 
\\
V(\epsilon,z)\sim \frac{\sqrt{\frac{2}{\pi}}e^{\frac{z^{2}}{4}}}{z^{-\epsilon+\frac{1}{2}}}
\left(1+\frac{\left(\epsilon-\frac{1}{2}\right)\left(\epsilon-\frac{3}{2}\right)}{2z^{2}}+\cdots \right). 
\end{gather*}
$V(\epsilon,x)$ diverge for $z \rightarrow \infty$ but $U(\epsilon,x)$ is an appropriate solution and we thus take
\begin{equation}
y_{\rm II}(\epsilon,z)=
\begin{cases}
U(\epsilon,z), &   z \geq b ,\\
0, &        z<b   .
\end{cases}
\end{equation}

\begin{figure}[t]
 \begin{minipage}[b]{.45\linewidth}\centering
\includegraphics[width=\linewidth]{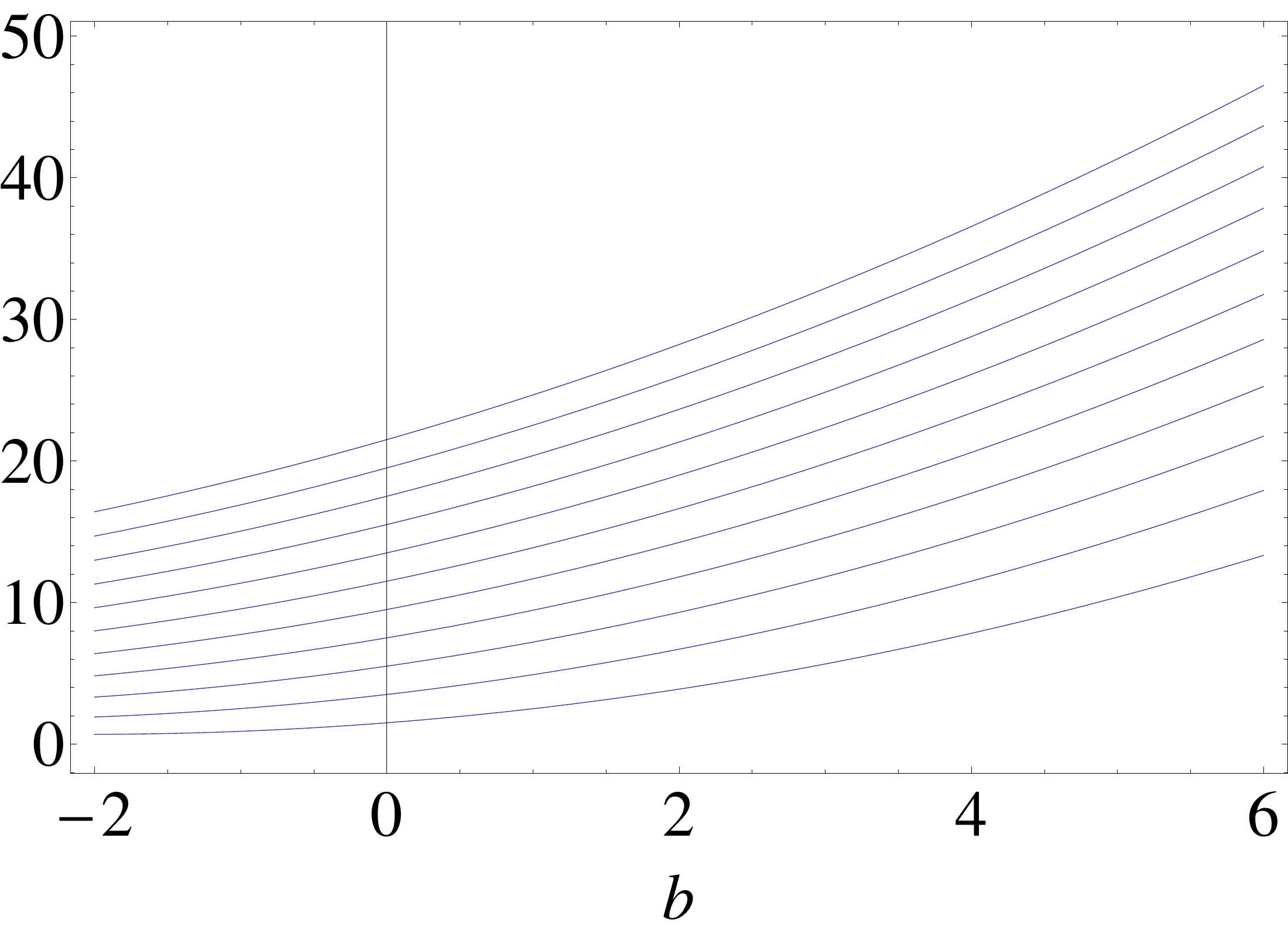}
 \caption{$E_{n}^{\rm II}(b)$ for the ground state and the 10 f\/irst excited states, $\hbar=1$ and $\omega=1$ for the Case $V_{\rm II}$.}\label{Fig1}
 \end{minipage} \hfill
 \begin{minipage}[b]{.45\linewidth}\centering
\includegraphics[width=\linewidth]{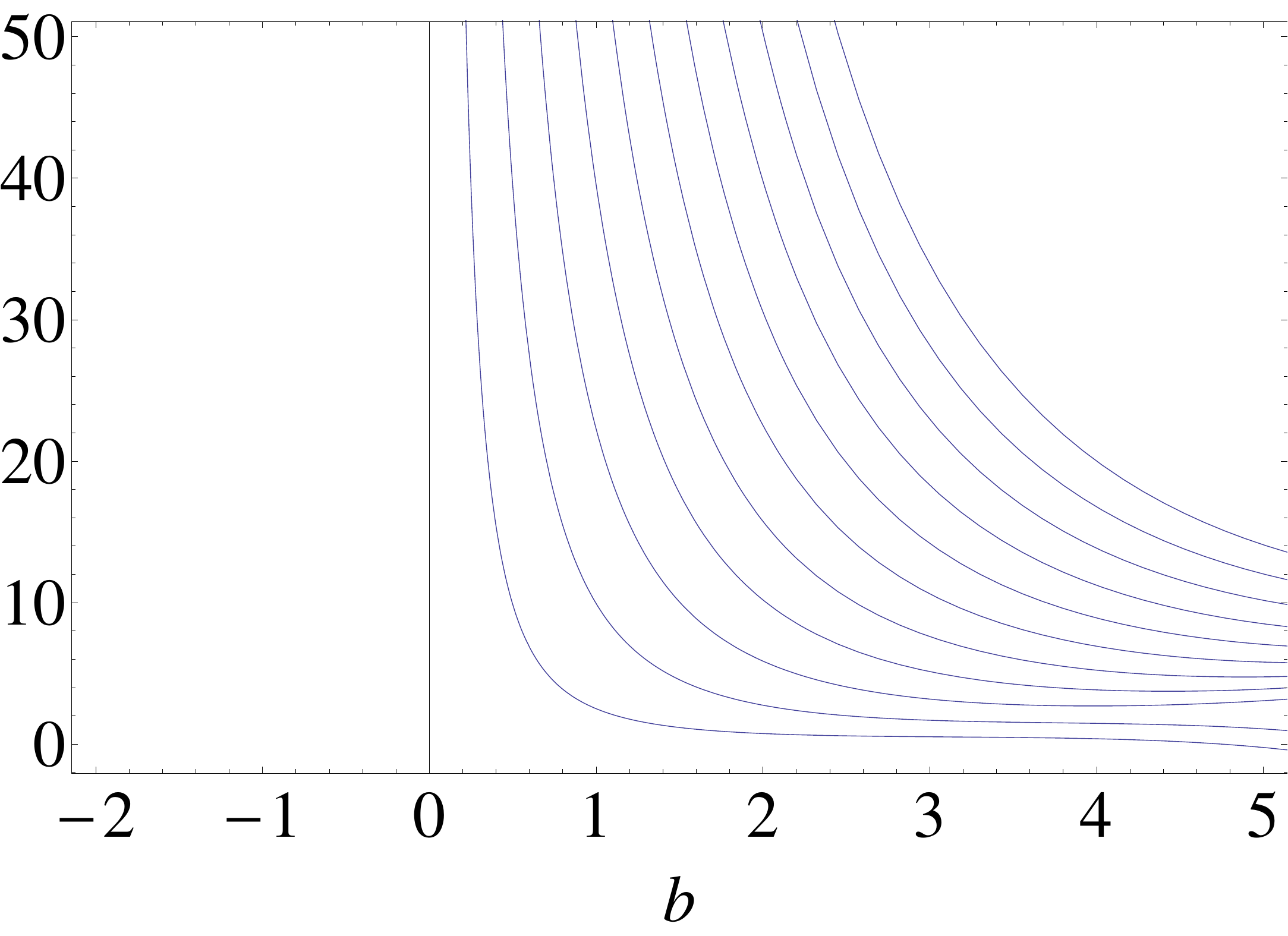}
 \caption{$E_{n}^{\rm III}(b)$  for the ground state and the 10 f\/irst excited states, $\hbar=1$ and $\omega=1$ for the Case $V_{\rm III}$.}\label{Fig3}
 \end{minipage}\hfill
\end{figure}

Let us present some results \cite{Dea} concerning the discrete ensemble of solutions. It can be shown (from $U(\epsilon_{n}^{\rm II}(b),b)=0$) that the energy levels depend of the position denoted $b$ of the inf\/inite barrier ( with $b$ assumed to be positive ) in the following way
\begin{equation*}
\epsilon_{n}^{\rm II}(b)=-\frac{1}{2}+\epsilon_{0}(n)+\epsilon_{1}(n)b+\epsilon_{2}(n)b^{2}+\cdots ,\qquad n=0,1,2,\dots,
\end{equation*}
and the f\/irst coef\/f\/icients are given by the following recurrence relations
\begin{gather*}
\epsilon_{0}(n)=-(2n+1),\qquad  n=0,1,2,\dots , \\
\epsilon_{1}(n+1)= \frac{2n+3}{2n+2} \epsilon_{1}(n),\qquad
\epsilon_{1}(0)=-2^{\frac{1}{2}}\pi^{-\frac{1}{2}},\\
\epsilon_{2}(n+1)=\left(\frac{2n+3}{2n+2}\right)^{2}\epsilon_{2}(n)+\frac{(2n+3)!(2n+2)!}{16\pi ( (n+1)!)^{4}(n+1) 2^{4n}},\\
\epsilon_{2}(0)=-2\pi^{-1}(1-\operatorname{Log}[2]).
\end{gather*}
These recurrence relations can be solved to obtain explicitly the value of the coef\/f\/icients. The coef\/f\/icients of higher-order terms can be calculated and this calculation could also be implemented numerically. Therefore, for every f\/ixed value of $b$ where the inf\/inite barrier is located, the zeros of this equation provide a discrete ensemble of solutions~\cite{Dea,Marq,Mei} $\epsilon_{n}^{\rm II}(b)$, $n=0,1,2,\dots $ and for the Hamiltonian with the potential given as in equation~\eqref{vII} we have $E_{n}^{\rm II}(b)=-\hbar\omega\epsilon_{n}^{\rm II}(b)$. The corresponding eigenfunctions are then $U(\epsilon_{n}^{\rm II}(b),z)$. The Fig.~\ref{Fig1} presents the energy for the ground state and the 10 f\/irst excited states as a function of the position of the inf\/inite barrier. The Fig.~\ref{Fig2} presents the ground state for an inf\/inite barrier located at $b=1$. The only exactly solvable case \cite{Dea,Marq,Mei} is when $b=0$. In this particular case we obtain from the condition $U(\epsilon_{n}^{\rm II},0)=0$
\begin{equation*}
E_{n}^{\rm II}=-\hbar\omega \epsilon_{n}=\hbar\omega\left(2n+1+\frac{1}{2}\right),\qquad n=0,1,2,\dots  .
\end{equation*}
The wavefunctions correspond to the odd levels of the harmonic oscillator. The limiting case where $b\rightarrow \infty$ allow to recover the spectrum of the harmonic oscillator given by  equation~\eqref{hamil}.

\subsection[Constrained harmonic oscillator $V_{\rm III}$]{Constrained harmonic oscillator $\boldsymbol{V_{\rm III}}$}

For the system $V_{\rm III}$ given by equation~\eqref{vIII}, we have the following constraints \cite{Dea, Marq}
\begin{enumerate}\itemsep=0pt
\item[$(i)$] $y(z)$ must be square integrable,

\item[$(ii)$] $y(z)$ must be continuous on $\mathbb{R}$, i.e.\ $y(z)\rightarrow 0$ when $ z \rightarrow b^{+}$ and $ z \rightarrow -b^{-}$.
\end{enumerate}

The associated eigenfunctions are the even and odd functions $y_{1}(\epsilon,z)$ and $y_{2}(\epsilon,z)$.
\begin{equation*}
y_{\rm III}(\epsilon,z)=\begin{cases}
y_{1}(\epsilon,z),  & n \  {\rm odd}, \quad  -b \leq z \leq b, \\
y_{2}(\epsilon,z),   & n \  {\rm even}, \quad  -b \leq z \leq b,\\
0,      &   z<-b ,\quad z>b  .
\end{cases}
\end{equation*}

\begin{figure}[t]
 \begin{minipage}[b]{.45\linewidth}\centering
\includegraphics[width=\linewidth]{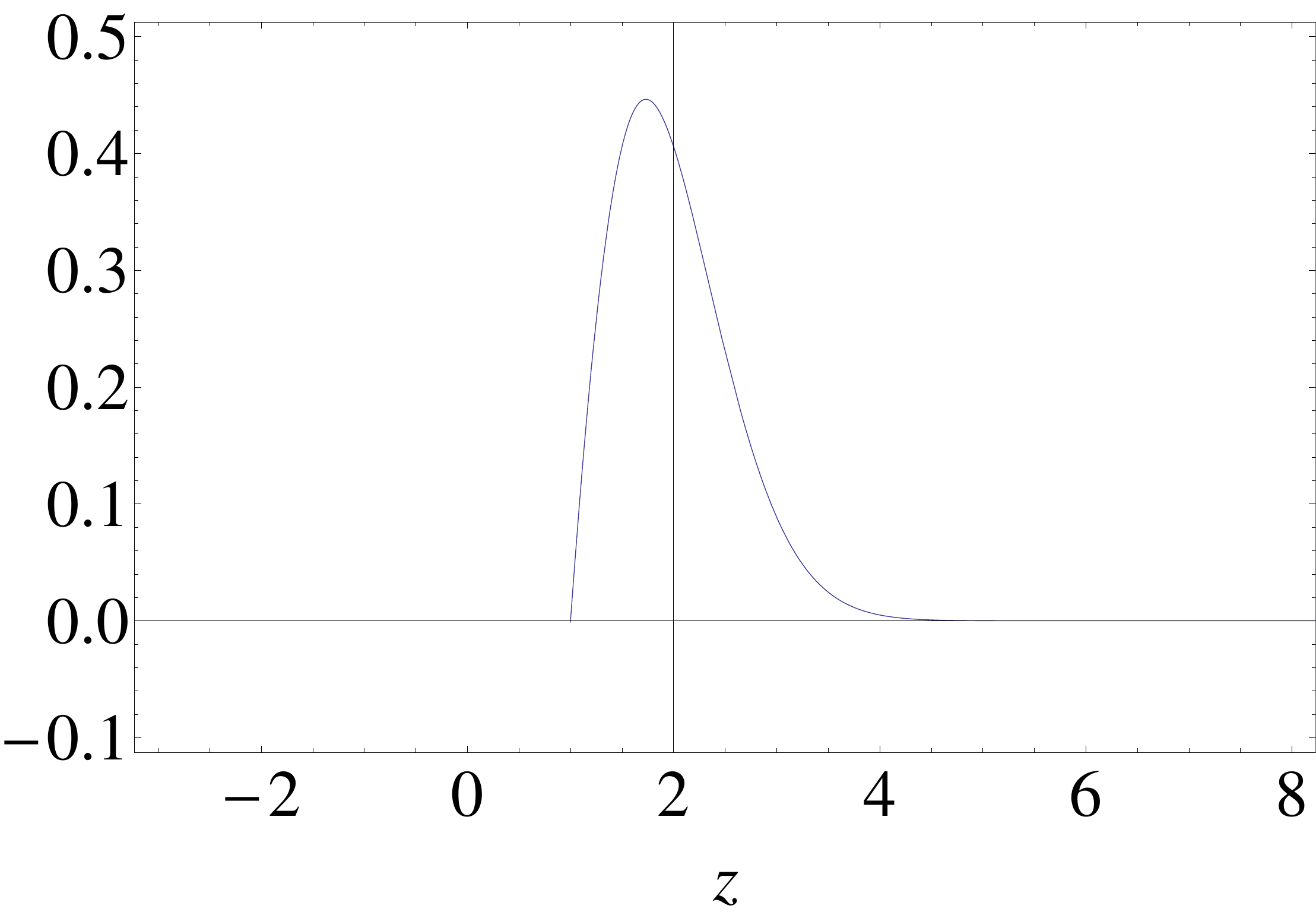}
 \caption{The wavefunction for the ground state $U(\epsilon_{0}^{\rm II},z)$ (with $\hbar=1$ and $b=1$) for the Case $V_{\rm II}$.}\label{Fig2}
 \end{minipage} \hfill
 \begin{minipage}[b]{.45\linewidth}\centering
\includegraphics[width=\linewidth]{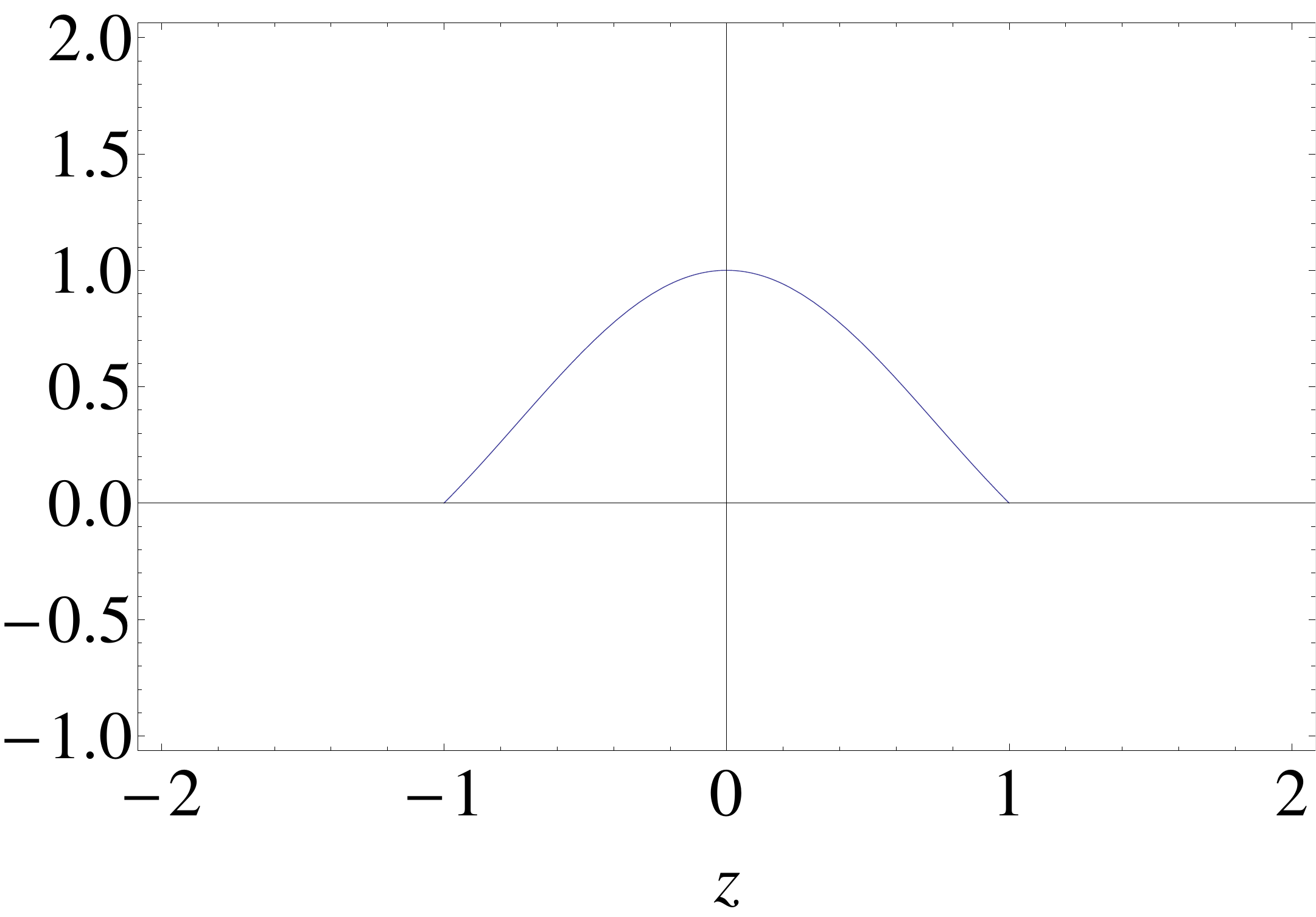}
 \caption{The wavefunction for the ground state $y_{1}(\epsilon_{0}^{\rm III},z)$ (with $\hbar=1$ and $b=1$) for the Case $V_{\rm III}$.}\label{Fig4}
 \end{minipage}\hfill
\end{figure}

The expansion of the eigenvalues as a function of the positions of the barriers takes the form in this case~\cite{Dea}
\begin{equation*}
\epsilon_{n}^{\rm III}(b)=\frac{\epsilon_{-2}(n)}{b^{2}}+\epsilon_{0}(n)+\epsilon_{2}(n)b^{2}+\epsilon_{4}(n)b^{4}+\epsilon_{6}(n)b^{6}+\cdots ,\qquad n=0,1,2,\dots,
\end{equation*}
with the f\/irst coef\/f\/icients given by
\begin{gather*}
\epsilon_{-2}(n)=-\left(\frac{n+1}{2}\right)^{2} \pi^{2},\qquad  n=0,1,2,\dots, \\
\epsilon_{0}(n)=\epsilon_{4}=0 ,\qquad
\epsilon_{2}(n)=-\frac{1}{12}-\frac{1}{8\epsilon_{-2}(n)},\\
\epsilon_{6}(n)=\frac{1}{720\epsilon_{-2}(n)}+\frac{5}{192\epsilon_{-2}(n)^{2}}+\frac{7}{128 \epsilon_{-2}(n)^{3}}.
\end{gather*}
These relations thus provide for a f\/ixed value of $b$ a discrete ensemble of solutions $\epsilon_{n}^{\rm III}(b)$, $n=0,1,2,\dots $.
For the system given by the equation~\eqref{vIII} the energy spectrum is thus $E_{n}^{\rm III}(b)=-\hbar\omega\epsilon_{n}^{\rm III}(b)$.
The corresponding eigenfunctions are then $y_{1}(\epsilon_{n}^{\rm III}(b),z)$ for the even levels and $y_{1}(\epsilon_{n}^{\rm III}(b),z)$ for the odd levels. The Fig.~\ref{Fig3} presents the energy for the ground state and the 10 f\/irst excited states as a function of the position of the inf\/inite barrier. The Fig.~\ref{Fig4} presents the ground state for an inf\/inite barrier located at $b=1$.

\section{Factorization method and singular Hamiltonian}\label{section3}

In supersymmetric quantum mechanics \cite{Gen,Jun,Wit}, we can def\/ine two f\/irst-order operators $A$ and $A^{\dagger}$ (supercharges) in the following way{\samepage
\begin{equation*}
A=\frac{\hbar}{\sqrt{2}}\frac{d}{dx}+W(x),\qquad A^{\dagger}=-\frac{\hbar}{\sqrt{2}}\frac{d}{dx}+W(x)  . 
\end{equation*}
where $W(x)$ is called the superpotential.}

We consider the following two Hamiltonians which are called ``superpartners''
\begin{equation}
H_{1}=A^{\dagger}A=-\frac{\hbar^{2}}{2}\frac{d^{2}}{dx^{2}}+W^{2}-\frac{\hbar}{\sqrt{2}}W',\qquad H_{2}= AA^{\dagger}=-\frac{\hbar^{2}}{2}\frac{d^{2}}{dx^{2}}+W^{2}+\frac{\hbar}{\sqrt{2}}W'. \label{h1h2}
\end{equation}
It can be shown that these algebraic relations allow to relate eigenfunctions and energy spectrum of the two superpartners $H_{1}$ and $H_{2}$. There are two cases that can be considered. The f\/irst one corresponds to broken supersymmetry and happen when $A\psi_{0}^{(1)}\neq 0$, $E_{0}^{(1)}\neq 0$, $A^{\dagger}\psi_{0}^{(2)}\neq 0$ and $E_{0}^{(2)}\neq 0$. The wavefunctions and the energy spectra are thus related in the following way
\begin{equation}
E_{n}^{(2)}=E_{n}^{(1)}>0,\qquad \psi_{n}^{(2)}=\big(E_{n}^{(1)}\big)^{-\frac{1}{2}}A\psi_{n}^{(1)},\qquad  \psi_{n}^{(1)}=\big(E_{n}^{(2)}\big)^{-\frac{1}{2}}A^{\dagger}\psi_{n}^{(2)}  , \label{enb1}
\end{equation}
and the two Hamiltonians are strictly isospectral.

For the second case the supersymmetry is unbroken and one of the Hamiltonians allows a~zero mode, i.e.\ a states is annihilated by one of the supercharges. Without lost of generality we take $H_{1}$ as having a zero mode (also referred as zero energy ground state) and thus we have $A\psi_{0}^{(1)}= 0$, $E_{0}^{(1)}= 0$, $A^{\dagger}\psi_{0}^{(2)}\neq 0$ and $E_{0}^{(2)}\neq 0$.  We obtain the following relations between the wavefunctions and the energy spectra
\begin{gather}
E_{n}^{(2)}=E_{n+1}^{(1)},\qquad E_{0}^{(1)}=0,\qquad \psi_{n}^{(2)}=\big(E_{n+1}^{(1)}\big)^{-\frac{1}{2}}A \psi_{n+1}^{(1)},\qquad  \psi_{n+1}^{(1)}=\big(E_{n}^{(2)}\big)^{-\frac{1}{2}}A^{\dagger} \psi_{n}^{(2)}  .\!\!\! \label{enb2}
\end{gather}
In this case, the superpotential $W(x)$ can be written in term $\psi_{0}^{(1)}$ as $W(x)= \frac{\psi_{0}^{(1)\prime}}{\psi_{0}^{(1)}}$.

These relations~\eqref{enb1} and~\eqref{enb2} allow, if one of these systems ($H_{1}$ or $H_{2}$) is exactly solvable, to obtain the energy spectrum and the wavefunctions of its superpartner. A such construction can be extended more generally to families of superpartners and superpartners of $k$th-order. It can be shown, that in the case of Hermitian operators, the factorization relations~\eqref{h1h2} are equivalent to the following intertwinning (also called Darboux) relations
\begin{equation}
H_{1}A=AH_{2}, \qquad A^{\dagger}H_{1}=H_{2}A^{\dagger}. \label{int}
\end{equation}
The factorization and interwining relations can be combined and written in terms of matrices that generate a superalgebra. These algebraic relations also allow to relate the creation and annihilation operators of the Hamiltonians $H_{1}$ and $H_{2}$. The ladder operators $c^{\dagger}$ and $c$ of $H_{1}$ and the ladder operators of $H_{2}$ that we denote $M^{\dagger}$ and $M$ are related by
\begin{equation}
M=A^{\dagger}cA,\qquad M^{\dagger}=A^{\dagger}c^{\dagger}A, \label{lad}
\end{equation}
However, as mentioned earlier the algebraic relations given by equations~\eqref{h1h2},~\eqref{int} and~\eqref{lad}
can be formal and do not take into account boundaries and singularities. In the case of a 1D nonsingular Hamiltonian~$H_{1}$ and a singular supersymmetric partner $H_{2}$ the wavefunctions $\psi^{(2)}$ obtained from those of $H_{1}$ (i.e.~$\psi^{(1)}$) and by acting with the supercharges (i.e.\ using~\eqref{enb1} or~\eqref{enb2}) on are not guaranteed to belong in the Hilbert space of~$H_{2}$. If the supercharges admit singularity the resulting wavefunctions could also admit singularity.

It is known that a singularity of the form $\frac{1}{(x-a)^{2}}$ makes that a particle is conf\/ined in one of the regions (i.e.\ left or right side of the singularity) because the probability of going through the barrier is zero. We thus need to compute independently the eigenfunctions for each region divided by a such singularity \cite{Fra,Lat}. It was shown by Marquez, Nieto and Negro~\cite{Marq} that for Hamiltonians with such singular terms we can construct an adequate superpartner by using inf\/inite barriers placed where are located the singularities on the real axis and recover isospectrality between the superpartners. The wavefunctions and the energy spectrum of the systems with singularities can now be obtained from those of the regular Hamiltonian with inf\/inite barriers and the equations~\eqref{enb1} and~\eqref{enb2}. Using this approach, families of quantum systems that are superpartners of the harmonic oscillator with one singularity were studied \cite{Marq}.

\section{Singular nonlinear oscillator}\label{section4}
Let us consider the following supercharges that admit singularities at $a$ and $-a$
\begin{gather*}
A^{\dagger}= \frac{ \hbar}{\sqrt{2}}\left(- \frac{d}{dx} - \frac{1}{2a^{2}}x
- \left(\frac{-1}{x-a}+\frac{-1}{x+a}\right)\right),  
\\
A= \frac{ \hbar}{\sqrt{2}}\left( \frac{d}{dx} - \frac{
1}{2a^{2}}x - \left(\frac{-1}{x-a}+\frac{-1}{x+a}\right)\right) . 
\end{gather*}
We can form from these operators, the following Hamiltonian $H_{1}$ and $H_{2}$
\begin{gather}
H_{1}=A^{\dagger}A
=\frac{P_{x}^{2}}{2}+V_{1}=\frac{P_{x}^{2}}{2}+\frac{\hbar^{2}x^{2}}{8a^{4}}
+\frac{\hbar^{2}}{(x-a)^{2}}+\frac{\hbar^{2}}{(x+a)^{2}}-\frac{3\hbar^{2}}{4a^{2}}, \label{singh1}
\\
H_{2}=AA^{\dagger}=\frac{P_{x}^{2}}{2}+V_{2}=\frac{P_{x}^{2}}{2}+\frac{\hbar^{2} x^{2}}{8 a^{4}}-
\frac{5\hbar^{2}}{4a^{2}}   . \label{singh2}
\end{gather}
These algebraic relations can be used in the case $a=ia_{0}$, $a_{0}\in \mathbb{R}$ to solve the nonsingular isotonic oscillator, but in the singular case $a \in \mathbb{R}$ they can not be used. The Hamiltonian $H_{2}$ is the well known harmonic oscillator with an additive constant and in principle the SUSYQM relations would provide the corresponding wavefunctions and energy spectrum of the Hamiltonian~$H_{1}$. However, the operators $A^{\dagger}$ contains singularities and using SUSYQM relations, we obtain eigenstates of the Hamiltonian $H_{1}$ but non square integrable states and thus non physical states. However, as mentioned this is possible to recover the isospectrality by using an appropriate superpartner for each region we would like to compute the wavefunctions and def\/ined by the singularities, i.e.\ $R_{1}=(-\infty,-a]$, $R_{2}=[a,\infty)$ and $R_{3}=[-a,a]$. Due to symmetry, the calculation for the f\/irst and the second regions are similar. Let us only present, the calculation in the case of the regions $R_{2}$ and $R_{3}$.

In the case of the singular Hamiltonian $H_{1}$ in the region $R_{2}$ we consider a modif\/ied superpartner that is the potential given by equation~\eqref{vII} up to an additive constant that does not af\/fect the results obtained in Section~\ref{section2} is
\begin{equation*}
V_{2_{\rm II}}(x)=
\begin{cases}
\dfrac{\hbar^{2} x^{2}}{8 a^{4}}-
\dfrac{5\hbar^{2}}{4a^{2}} , &  x \geq a,\quad a>0 ,  \\
\infty, &  x < a, \quad a \in \mathbb{R}.
\end{cases}
\end{equation*}

In the case of the singular Hamiltonian $H_{1}$ in the region $R_{3}$ we take a modif\/ied superpartner as given by equation~\eqref{vII} up to an additive constant
\begin{equation*} 
V_{2_{\rm III}}(x)=
\begin{cases}
\dfrac{\hbar^{2} x^{2}}{8 a^{4}}-
\dfrac{5\hbar^{2}}{4a^{2}} , &   -a \leq x \leq a, \quad a>0 , \\
\infty, &  x < - a,  \quad x > a ,\quad a \in \mathbb{R}.
\end{cases}
\end{equation*}

The case with two singularities dif\/fers from the case with one singularity because of the possibility of bounded intervals as $R_{3}$ and as we saw from the Fig.~\ref{Fig1}--\ref{Fig4} the behavior of the two types of constrained harmonic oscillator dif\/fer and thus their superpartners will also admit dif\/ferent spectrum and wavefunctions. We can now apply results of Sections~\ref{section2} and~\ref{section4} and we obtain for the Hamiltonian given by equation~\eqref{singh1}.

Case $R_{2}=[a,\infty)$:
\begin{gather*}
\psi\big(\epsilon_{n}^{\rm II}(a),x\big)=A^{\dagger}y_{\rm II}(x)\\
\hphantom{\psi\big(\epsilon_{n}^{\rm II}(a),x\big)}{}
=\begin{cases}
\left[ \dfrac{\hbar}{\sqrt{2}}\left(- \dfrac{d}{dx} - \dfrac{1}{2a^{2}}x- \left(\dfrac{-1}{x-a}+\dfrac{-1}{x+a}\right)
\right) \right]U\big(\epsilon_{n}^{\rm II}(a),x\big),  & x \geq a,  \\
0,   &   x \leq  a,
\end{cases}
\\
E_{n}^{\rm II}=\frac{\hbar^{2}}{2a^{2}}\left(\epsilon_{n}^{\rm II}(a)-\frac{5}{2}\right). 
\end{gather*}

Case $R_{3}=[-a,a]$:
\begin{gather*}
\psi\big(\epsilon_{n}^{\rm III}(a),x\big)=A^{\dagger}y_{\rm III}(x)\\
=\begin{cases}
\left[ \dfrac{\hbar}{\sqrt{2}}\left(- \dfrac{d}{dx} - \dfrac{1}{2a^{2}}x- \left(\dfrac{-1}{x-a}+\dfrac{-1}{x+a}\right)
\right) \right]y_{1}\big(\epsilon_{n}^{\rm III}(a),x\big),  &n \  {\rm even}, \ x \leq |a|,  \vspace{1mm}\\
\left[ \dfrac{\hbar}{\sqrt{2}}\left(- \dfrac{d}{dx} - \dfrac{1}{2a^{2}}x- \left(\dfrac{-1}{x-a}+\dfrac{-1}{x+a}\right)
\right) \right]y_{2}\big(\epsilon_{n}^{\rm III}(a),x\big),  & n \  {\rm odd}, \ x \leq |a|,  \\
0,   &   x \geq  |a|,\\
\end{cases}
\\
E_{n}^{\rm III}=\frac{\hbar^{2}}{2a^{2}}\left(\epsilon_{n}^{\rm III}(a)-\frac{5}{2}\right). 
\end{gather*}
The wavefunctions are square integrable and also continuous at the inf\/inite barrier by construction. This can be obtained by considering the local behavior in terms of a Taylor series about the position of the inf\/inite barrier. The completeness of the set is also provided by the fact that the state annihilated by the operator~$A$ has singularities at the inf\/inite barrier and is not an admissible wavefunction. In the next section we will show how these results can also be used in context of superintegrable systems that also involve such singular systems.

\section{SUSYQM and superintegrability}\label{section5}

Let us now consider the following 2D superintegrable Hamiltonian
\begin{equation}
H_{s1}=H_{x_{1}}+H_{x_{2}} = \frac{1}{2}P_{x_{1}}^{2}+\frac{1}{2}P_{x_{2}}^{2}+\hbar^{2}\left( \frac{x_{1}^{2}+x_{2}^{2}}{8a^{4}} +
\frac{1}{(x_{1}-a)^{2}}+\frac{1}{(x_{1}+a)^{2}}\right)  . \label{superh}
\end{equation}

This system is one of the four irreducible superintegrable systems with a second and a third-order integrals of motion found by Gravel \cite{Gra} that remained to be solved. An important property of superintegrable systems with higher-order integrals of motion is that
the quantum systems and their classical analog do not necessarily coincide~\cite{Gra}. In this case the classical limit is the free particle. It involves the nonsingular or the singular isotonic oscillator as we choose $a=ia_{0}$, $a_{0}\in \mathbb{R}$ or  $a \in \mathbb{R}$. The existence of the integrals of motion does not depend on the nature of the singularities and thus of this choice. This system was studied \cite{Mar4, Mar3} using polynomial algebras for both cases. However, only in the nonsingular case the f\/inite-dimensional unitary representations correspond to physical states. In the singular case, the integrals (i.e.\ they commute with the Hamiltonian $H_{s1}$) $I_{1}$ and $I_{2}$ take the form
\begin{gather*}
I_{1} = P_{x_{1}}^{2} - P_{x_{2}}^{2} + 2\hbar^{2}\left( \frac{x_{1}^{2}-x_{2}^{2}}{8a^{4}} +
\frac{1}{(x_{1}-a)^{2}}+\frac{1}{(x_{1}+a)^{2}}\right),
\\
I_{2} = \frac{1}{2}\big\{L,P_{x_{1}}^{2}\big\} + \frac{1}{2}\hbar^{2}\left\{x_{2}\left(
\frac{4a^{2}-x_{1}^{2}}{4a^{4}} -
\frac{6(x_{1}^{2}+a^{2})}{(x_{1}^{2}-a^{2})^{2} }\right),P_{x_{1}}\right\}
\\
\hphantom{I_{2} =}{} + \frac{1}{2}\hbar^{2}\left\{x_{1}\left(\frac{(x_{1}^{2}-4a^{2})}{4a^{4}} -
\frac{2}{x_{1}^{2}-a^{2}} + \frac{4(x_{1}^{2}+a^{2})}{(x_{1}^{2}-a^{2})^{2}}
\right),P_{x_{2}}\right\},
\end{gather*}
where $\{ \; , \; \}$ is the anticommutator. They are respectively polynomials in the momenta of order~2 and~3. These integrals generate the following commutation relations~\cite{Mar4, Mar3}
\begin{gather*}
[I_{1},I_{2}]=I_{3} , \qquad [I_{1},I_{3}]=\frac{4h^{4}}{a^{4}}I_{2}, \\
 [I_{2},I_{3}]= -2\hbar^{2}I_{1}^{3} - 6\hbar^{2}I_{1}^{2}H_{s1} + 8\hbar^{2}H_{s1}^{3}+ 6\frac{\hbar^{4}}{a^{2}}I_{1}^{2}\\
\hphantom{[I_{2},I_{3}]=}{}
+ 8\frac{\hbar^{4}}{a^{2}}H_{s1}I_{1}-8\frac{\hbar^{4}}{a^{2}}H_{s1}^{2} + 2\frac{\hbar^{6}}{a^{4}}I_{1} - 2\frac{\hbar^{6}}{a^{4}}H_{s1} -
6\frac{\hbar^{8}}{a^{6}}.
\end{gather*}

We can calculate the Casimir operator, the realizations in terms of deformed oscillator al\-gebras, the f\/inite-dimensional unitary representation and the corresponding energy spectrum. For $a \in \mathbb{R}$ the following solutions is obtained
\begin{equation*}
E=\frac{\hbar^{2}(p+3)}{2a^{2}},\qquad
\Phi(N)= \frac{\hbar^{8}}{a^{4}} N(p+1-N)(N+1)(N+3), 
\end{equation*}
where $\Phi(N)$ is the structure functions of the deformed oscillator algebra. However, the polynomial algebra does not take into account boundaries and singularities and is thus only a formal algebraic construction in this case and this f\/inite-dimensional unitary representations do not correspond to physical states. Furthermore, this is the same energy spectrum that is obtained by using the formal factorization and Darboux relations with the harmonic oscillator and given by equation~\eqref{singh1} and~\eqref{singh2}. However, as mentioned we can show they are not square integrable.

The fact that the polynomial algebra and the SUSYQM relation are formal is also related. This can be understood from the fact that the integrals and the polynomial algebras can be derived using SUSYQM~\cite{Mar2}. In the $x_{1}$-axis, the ladder operators are constructed from supersymmetry with the supercharges using equation~\eqref{lad}~\cite{Mar2,Mar4}. The annihilation operators are given by
\begin{gather*}
M_{x_{1}}=
\frac{\hbar^{2}}{4a^{2}}\left(-\frac{d}{dx_{1}} - \frac{1}{2a^{2}}x_{1}
+\left(\frac{1}{x_{1}-a}+\frac{1}{x_{1}+a}\right)\right)\\
\phantom{M_{x_{1}}=}{}
\times \left(x_{1}+2a^{2}\frac{d}{dx_{1}}\right)\left(\frac{d}{dx_{1}} - \frac{1}{2a^{2}}x_{1} +\left(\frac{1}{x_{1}-a}+\frac{1}{x_{1}+a}\right)\right), 
\\
 M_{x_{2}}=\frac{\hbar}{2a^{2}}\left(x_{2}+2a^{2}\frac{d}{dx_{2}}\right)  , 
\end{gather*}
with the creation operators obtained by considering ($M_{x_{1}})^{\dagger}$ and  $(M_{x_{2}})^{\dagger}$. The integrals can be written as
\begin{equation*}
I_{2}=\frac{-2a^{2}i}{\hbar}\big(M_{x_{1}}^{\dagger}M_{x_{2}}- M_{x_{1}}M_{x_{2}}^{\dagger}\big), \qquad I_{3}=\frac{-2a^{2}i}{\hbar}
\big(M_{x_{1}}^{\dagger }M_{x_{2}}+ M_{x_{1}}M_{x_{2}}^{\dagger}\big)  . 
\end{equation*}

\subsection{Modif\/ied SUSYQM and singular superintegrable systems}

Let us consider again the Hamiltonian $H_{s1}$ given by the equation~\eqref{superh} and use the results of Section~\ref{section4} on
the 1D Hamiltonian $H_{x_{1}}$ and takes a modif\/ied superpartner that is isospectral using inf\/inite barrier at the singularities. The 1D Hamiltonian $H_{x_{2}}$ is only the 1D unconstrained harmonic oscillator.

The wavefunctions and the energy spectrum would be thus for a particle in the region $[a,\infty)$ for the axis $x_{1}$
\begin{gather*}
 \Phi(x_{1},x_{2})=\chi(x_{2})A^{\dagger}y_{\rm II}^{(a)}(x_{1})
\\
=\begin{cases}
\left[ \chi(x_{2})\dfrac{\hbar}{\sqrt{2}}\left(- \dfrac{d}{dx_{1}} - \dfrac{1}{2a^{2}}x_{1}-
\left(\dfrac{-1}{x_{1}-a}+\dfrac{-1}{x_{1}+a}\right)\right) \right]U\big(\epsilon_{n}(a),x_{1}\big),  & x_{1} \geq a,  \\
0,   &   x_{1} \leq  a,
\end{cases}
\\
E=\frac{\hbar^{2}}{2a^{2}}\big(\epsilon_{n}^{\rm II}(a)+k-2\big). 
\end{gather*}

For a particle in the region $[-a,a]$ for the axis $x_{1}$
\begin{gather*} 
\Phi(x_{1},x_{2})=\chi(x_{2})A^{\dagger}y_{\rm III}^{(a)}(x_{1})
\\
=\begin{cases}
\left[ \chi(x_{2})\dfrac{\hbar}{\sqrt{2}}\left(- \dfrac{d}{dx_{1}} - \dfrac{1}{2a^{2}}x_{1}-
\left(\dfrac{-1}{x_{1}-a}+\dfrac{-1}{x_{1}+a}\right)\right) \right]y_{1}(\epsilon_{n}(a),x_{1}),\!  & n \ {\rm even}, \  x_{1} \leq |a|,
\vspace{1mm} \\
\left[ \chi(x_{2})\dfrac{\hbar}{\sqrt{2}}\left(- \dfrac{d}{dx_{1}} - \dfrac{1}{2a^{2}}x_{1}- \left(\dfrac{-1}{x_{1}-a}+\frac{-1}{x_{1}+a}\right)\right) \right]y_{2}(\epsilon_{n}(a),x_{1}),\!  & n \ {\rm odd}, \ x_{1} \leq |a|,  \\
0,   &   x_{1} \geq  |a|,\\
\end{cases}
\\
E=\frac{\hbar^{2}}{2a^{2}}\big(\epsilon_{n}^{\rm III}(a)+k-2\big), 
\end{gather*}
with
\begin{equation*}
\chi(x_{2})=Ce^{\frac{-1}{4a^{2}}x_{2}^{2}}H_{k}\left(\sqrt{\frac{1}{2a^{2}}}x_{2}\right). 
\end{equation*}

The eigenfunctions and energy spectrum for the three other irreducible 2D superintegrable systems with a second and third-order integrals of motion can be calculated using the same approach.

\textbf{Hamiltonian $H_{s2}$}
\begin{equation*}
H_{s2}= \frac{1}{2}P_{x_{1}}^{2}+\frac{1}{2}P_{x_{2}}^{2}+\hbar^{2}\left( \frac{x_{1}^{2}+9 x_{2}^{2}}{8a^{4}} +
\frac{1}{(x_{1}-a)^{2}}+\frac{1}{(x_{1}+a)^{2}}\right)   . 
\end{equation*}

For a particle in the region $[a,\infty)$ for the axis $x_{1}$
\begin{gather*}
 \Phi(x_{1},x_{2})=\chi(x_{2})A^{\dagger}y_{\rm II}^{(a)}(x_{1})
\\
=\begin{cases}
\left[ \chi(x_{2})\dfrac{\hbar}{\sqrt{2}}
\left(- \dfrac{d}{dx_{1}} - \dfrac{1}{2a^{2}}x_{1}- \left(\dfrac{-1}{x_{1}-a}+\dfrac{-1}{x_{1}+a}\right)
\right) \right]U(\epsilon_{n}(a),x_{1}),  & x_{1} \geq a,  \\
0,   &   x_{1} \leq  a,
\end{cases}
\\
E=\frac{\hbar^{2}}{2a^{2}}\big(\epsilon_{n}^{\rm II}(a)+k-1\big).
\end{gather*}

For a particle in the region $[-a,a]$ for the axis $x_{1}$
\begin{gather*}
 \Phi(x_{1},x_{2})=\chi(x_{2})A^{\dagger}y_{\rm III}^{(a)}(x_{1})
\\
=\begin{cases}
\left[ \chi(x_{2})\dfrac{\hbar}{\sqrt{2}}
\left(- \dfrac{d}{dx_{1}} - \dfrac{1}{2a^{2}}x_{1}-
\left(\dfrac{-1}{x_{1}-a}+\dfrac{-1}{x_{1}+a}\right)\right) \right]y_{1}(\epsilon_{n}(a),x_{1}),\!   & n \ {\rm even}, \ x_{1} \leq |a|,
\vspace{1mm} \\
\left[ \chi(x_{2})\dfrac{\hbar}{\sqrt{2}}\left(- \dfrac{d}{dx_{1}} - \dfrac{1}{2a^{2}}x_{1}- \left(\dfrac{-1}{x_{1}-a}+\dfrac{-1}{x_{1}+a}\right)\right) \right]y_{2}(\epsilon_{n}(a),x_{1}),\!   & n \ {\rm odd}, \ x_{1} \leq |a|,  \\
0,   &   x_{1} \geq  |a|,
\end{cases}
\\
E=\frac{\hbar^{2}}{2a^{2}}\big(\epsilon_{n}^{\rm III}(a)+k-1\big), 
\end{gather*}
with
\begin{equation*}
\chi(x_{2})=Ce^{\frac{-3}{4a^{2}}x_{2}^{2}}H_{k}\left(\sqrt{\frac{3}{2a^{2}}}x_{2}\right). 
\end{equation*}

\textbf{Hamiltonian $H_{s3}$}
\begin{equation*}
H_{s3}= \frac{1}{2}P_{x_{1}}^{2}+\frac{1}{2}P_{x_{2}}^{2}+\hbar^{2}\left( \frac{x_{1}^{2}+x_{2}^{2}}{8a^{4}} +\frac{1}{x_{2}^{2}}+
\frac{1}{(x_{1}-a)^{2}}+\frac{1}{(x_{1}+a)^{2}}\right)   . 
\end{equation*}

For a particle in the region $[a,\infty)$ for the axis $x_{1}$
\begin{gather*} 
\Phi(x_{1},x_{2})=\chi(x_{2})A^{\dagger}y_{\rm II}^{(a)}(x_{1})
\\
=\begin{cases}
\left[ \chi(x_{2})\dfrac{\hbar}{\sqrt{2}}\left(- \dfrac{d}{dx_{1}} - \dfrac{1}{2a^{2}}x_{1}- \left(\dfrac{-1}{x_{1}-a}+\dfrac{-1}{x_{1}+a}\right)\right) \right]U(\epsilon_{n}(a),x_{1}),  & x_{1} \geq a,  \\
0,   &   x_{1} \leq  a,
\end{cases}
\\
E=\frac{\hbar^{2}}{2a^{2}}\big(\epsilon_{n}^{\rm II}(a)+k\big). 
\end{gather*}

For a particle in the region $[-a,a]$ for the axis $x_{1}$
\begin{gather*} 
\Phi(x_{1},x_{2})=\chi(x_{2})A^{\dagger}y_{\rm III}^{(a)}(x_{1})
\\
=\begin{cases}
\left[ \chi(x_{2})\dfrac{\hbar}{\sqrt{2}}\left(- \dfrac{d}{dx_{1}} - \dfrac{1}{2a^{2}}x_{1}- \left(\dfrac{-1}{x_{1}-a}+\dfrac{-1}{x_{1}+a}\right)\right) \right]y_{1}(\epsilon_{n}(a),x_{1}),\!  & n \  {\rm even},  \ x_{1} \leq |a|,  \vspace{1mm}\\
\left[ \chi(x_{2})\dfrac{\hbar}{\sqrt{2}}\left(- \dfrac{d}{dx_{1}} - \dfrac{1}{2a^{2}}x_{1}- \left(\dfrac{-1}{x_{1}-a}+\dfrac{-1}{x_{1}+a}\right)\right) \right]y_{2}(\epsilon_{n}(a),x_{1}),\!  & n \ {\rm odd}, \ x_{1} \leq |a|,  \\
0,   &   x_{1} \geq  |a|,
\end{cases}
\\
E=\frac{\hbar^{2}}{2a^{2}}\big(\epsilon_{n}^{\rm III}(a)+k\big), 
\end{gather*}
with
\begin{equation*}
\chi(x_{2})=e^{-\frac{1}{4a^{2}}x_{2}^{2}}x_{2}^{2}L_{k}^{\frac{3}{2}}\left(\frac{1}{2a^{2}}x_{2}^{2}\right), 
\end{equation*}
where $L_{k}^{\frac{3}{2}}$ are Laguerre polynomials.

\textbf{Hamiltonian $H_{s4}$}
\begin{equation*}
H_{s4}= \frac{1}{2}P_{x_{1}}^{2}+\frac{1}{2}P_{x_{2}}^{2}+\hbar^{2}\left( \frac{x_{1}^{2}+ x_{2}^{2}}{8a^{4}} +
\frac{1}{(x_{1}-a)^{2}}+\frac{1}{(x_{1}+a)^{2}})+
\frac{1}{(x_{2}-a)^{2}}+\frac{1}{(x_{2}+a)^{2}}\right)   . 
\end{equation*}

For a particle in the region $[a,\infty)$ for the axis $x_{1}$ and $x_{2}$
\begin{gather*} \Phi(x_{1},x_{2})=\prod_{i}^{2} A_{i}^{\dagger}y_{\rm II}^{(a)}(x_{i}),
\end{gather*}
with
\begin{gather*} 
A_{i}^{\dagger}y_{\rm II}^{(a)}(x_{i})
=\begin{cases}
\left[ \dfrac{\hbar}{\sqrt{2}}\left(- \dfrac{d}{dx_{i}} - \dfrac{1}{2a^{2}}x_{i}-
\left(\dfrac{-1}{x_{i}-a}+\dfrac{-1}{x_{i}+a}\right)\right) \right]U(\epsilon_{n}(a),x_{i}),  & x_{i} \geq a,  \\
0,   &   x_{i} \leq  a,
\end{cases}
\\
E=\frac{\hbar^{2}}{2a^{2}}\big(\epsilon_{n}^{\rm II}(a)+k-5\big). 
\end{gather*}

For a particle in the region $[-a,a]$ for the axes $x_{1}$ and $x_2$
\begin{gather*} \Phi(x_{1},x_{2})= \prod_{i}^{2} A_{i}^{\dagger}y_{\rm III}^{(a)}(x_{i}),
\end{gather*}
with
\begin{gather*} 
A_{i}^{\dagger}y_{\rm III}^{(a)}(x_{i}) \\
=\begin{cases}
\left[ \dfrac{\hbar}{\sqrt{2}}\left(- \dfrac{d}{dx_{i}} - \dfrac{1}{2a^{2}}x_{i}-
\left(\dfrac{-1}{x_{i}-a}+\frac{-1}{x_{i}+a}\right)\right) \right]y_{1}(\epsilon_{n}(a),x_{i}),  & n \ {\rm even}, \ x_{i} \leq |a|,
\vspace{1mm}\\
\left[ \dfrac{\hbar}{\sqrt{2}}\left(- \dfrac{d}{dx_{i}} - \dfrac{1}{2a^{2}}x_{i}-
\left(\dfrac{-1}{x_{i}-a}+\dfrac{-1}{x_{i}+a}\right)\right) \right]y_{2}(\epsilon_{n}(a),x_{i}),  & n \  {\rm odd}, \ x_{i} \leq |a|,  \\
0,   &   x_{1} \geq  |a|,
\end{cases}
\\
E=\frac{\hbar^{2}}{2a^{2}}\big(\epsilon_{n}^{\rm III}(a)+k-5\big), 
\end{gather*}

\section{Conclusion}\label{section6}

In this paper, we studied the singular isotonic oscillator using the approach introduced in~\cite{Marq}. We obtained the energy spectrum and the wavefunctions from supersymmetry by identifying the appropriate superpartner. We also studied four singular 2D superintegrable Hamiltonians obtained by Gravel. The energy spectrum in the 1D cases is no longer equidistant and in the 2D superintegrable versions the energy spectra do not display accidental degeneracy explained by a~polynomial algebra that is only formal. However, this point out how supersymmetric quantum mechanics can be used to solve such systems. The wavefunctions are also interesting from the point of view of special functions as they are written in terms of expressions involving parabolic cylinder functions. These results could also be interesting in regard of possible applications. The constrained harmonic oscillators has found applications in the context of condensed matter~\cite{Fer}.

Moreover, the factorization is not unique \cite{Mie} and it was shown that the nonsingular isotonic oscillator has more general families of superpartners \cite{Mar5}. Many families of nonsingular superpartners of the radial oscillator were obtained and the relations with exceptional orthogonal polynomial studied \cite{Que}. The study of the singular version of these systems and the relations with special functions is also important. The approach described in this paper could also be applied to systems involving the fourth and the f\/ifth Painlev\'e transcendents that generate families of systems with singularities on the real axis for specif\/ic parameters \cite{And,Dav2,Dav1,Mar6,Mar2,Mar1}.

These results are also important as supersymmetry plays a role in the construction of new superintegrable systems in Cartesian \cite{Mar6, Mar5} but also in polar coordinates \cite{Dem,Pos}.

\subsection*{Acknowledgements}

{\sloppy This work was supported by the Australian Research Council through Discovery Project DP110101414. The article was written in part while he was visiting the Universite Libres de Bruxelles. He thanks C.~Quesne for her hospitality. He thanks the FNRS for a travel fellowship.

}

\pdfbookmark[1]{References}{ref}
\LastPageEnding

\end{document}